\documentclass{ws-procs9x6}

\def\beq{\begin{equation}}
\def\eeq{\end{equation}}
\def\beqa{\begin{eqnarray}}
\def\eeqa{\end{eqnarray}}

\def\za{\alpha}
\def\zb{\beta}
\def\lsim{\mathrel{\raise.3ex\hbox{$<$\kern-.75em\lower1ex\hbox{$\sim$}}} }
\def\gsim{\mathrel{\raise.3ex\hbox{$>$\kern-.75em\lower1ex\hbox{$\sim$}}} }

\begin{document}
%% net cover
%\addtolength{\baselineskip}{.2cm}
\thispagestyle{empty}

\onecolumn

\begin{flushright}
NCU-HEP-k008  \\
{KEK-TH-873} \\
May 2003
\end{flushright}

\vspace*{.5in}

\begin{center}
{\bf  \Large Neutrino Masses and Beyond from Supersymmetry }\\
\vspace*{.5in}
{\bf  Otto C.W. Kong}\\[.05in]
{\it Department of Physics, National Central University, Chung-li, TAIWAN 32054 \\
Theory Group, KEK, Tsukuba, Ibaraki, 305-0801, Japan \\
E-mail: otto@phy.ncu.edu.tw}

\vspace*{.8in}
{Abstract}\\
\end{center}
%\twocolumn[\maketitle\abstract{
%}]
A generic form of the supersymmetric SM naturally gives rise 
to the lepton number violating neutrino masses and mixings, without
the need for extra superfields beyond the minimal spectrum. 
Hence, SUSY can be consider the origin of beyond SM 
properties of neutrinos. We have developed a formulation 
under which one can efficiently analyze the model. 
Various sources of neutrino masses are discussed in details. 
Such mass contributions come from
lepton number and flavor violating couplings that
also give rise to a rich phenomenology of the neutrinos 
and other leptons, to be discussed.

\vfill
\noindent --------------- \\
$^\star$ Talk presented  at NOON 2003 (Feb 10-14), Kanazawa, Japan\\
 --- submission for the proceedings.  
 
\clearpage
\addtocounter{page}{-1}

\title{ Neutrino Masses and Beyond from Supersymmetry}

\author{OTTO~C.~W. KONG\footnote{\uppercase{W}ork partially
supported by grant \uppercase{NSC}91-2112-\uppercase{M}-008-042 of the 
\uppercase{N}ational \uppercase{S}cience \uppercase{C}ouncil of \uppercase{T}aiwan.}}

\address{Department of Physics, National Central University \\
Chung-li, TAIWAN 32054\\ 
\&\\  Theory Group, KEK,
Tsukuba, Ibaraki, 305-0801, Japan\\ 
E-mail:  otto@phy.ncu.edu.tw}

\maketitle

\abstracts{
A generic form of the supersymmetric SM naturally gives rise 
to the lepton number violating neutrino masses and mixings, without
the need for extra superfields beyond the minimal spectrum. 
Hence, SUSY can be considered the origin of beyond SM 
properties of neutrinos. We have developed a formulation 
under which one can efficiently analyze the model. Various sources of 
neutrino masses are discussed. Such mass contributions come from
lepton number and flavor violating couplings that also give rise to a rich 
phenomenology of the neutrinos and other leptons, also to be discussed.
}

\section{Introduction}
From the theoretical point of view, low-energy supersymmetry (SUSY) is by far
the most popular candidate theory for physics beyond the Standard Model (SM). 
On the experimental size, we now do have results confirming beyond SM
properties of neutrinos, which at least includes oscillations among different
neutrino spieces. The most natural way to have neutrino oscillations is to
have massive neutrinos. Here, we are particularly interested in properties of
such massive neutrinos that could actually be considered as arising from
SUSY. 

Within the SM, neutrino mass terms may be  described by VEVs of
dimension five operators of the form
\[
L_i \, \frac{\langle H \rangle
\langle H \rangle}{M} \, L_j \; ,
\]
where $M$ denotes some high energy scale. The non-renormalizable
dimension five operators should be considered as obtained from integrating
out some beyond SM physics underlying, physics of which can be probed
only at scale beyond $M$. Neutrino masses are usually classified as
Dirac or Majorana. Dirac mass terms involve  singlet fermions usually
named right-hand neutrinos ($\nu_{\scriptscriptstyle R}$ or 
$\nu_{\scriptscriptstyle S}$) giving rise to terms of the form
\[
\bar{\nu}_{\scriptscriptstyle S_k}
\langle H \rangle L_j \;.
\]
Lepton number violating {\it Majorana} mass terms at scale $M$
are typically introduced. Integrating out the heavy neutrino degrees
of freedom leaves the seesaw induced effective SM neutrino 
{\it Majorana} masses. Direct introduction of such Majorana masses without
heavy neutrino degrees of freedom have also been considered. The simplest
way to do that is to introduce a Higgs triplet with VEV, giving rise to
\[
L_i \langle T \rangle L_j \;.
\]
An effective triplet VEV $\langle T \rangle = \frac{\langle H \rangle
\langle H \rangle}{M}$ is always needed, though it can also come from
a loop diagram. Typically, we do need extra scalar bosons in the theory one
way or another\cite{zee}.
 
So, the old question of whether neutrinos are Dirac or Majorana is not quite
the right question to ask. Experimentally speaking, a (physical) neutrino is 
just a light  neutral fermion that experiences weak interactions. 
As long as low energy phenomenology is concerned, we only need to know
how many neutral fermion degrees of freedom are within reach, and what is the
generic mass matrix.

\section{Supersymmetry and Neutrinos}
A supersymmetric extension of the SM has four extra neutral fermions apart
from the SM ones. And nonzero masses are generally admissible for the
full set of seven neutral fermions including the neutrinos.

From the early history of supersymmetry (SUSY), there had been thinking about
its usage in the obviously non-supersymmetric low-energy phenomenology. One of 
the first idea was the identification of the neutrino as a goldstino, {\it i.e.} 
the Goldstone mode from (global) SUSY breaking\cite{AV}. Nowadays, the 
question : ``Is the masslessness of the neutrino a result of SUSY (breaking)?"
is obvious an uninteresting one. Nevertheless, {neutrinos} and {SUSY} just
may have everything to do with one another; after all, {\it nonzero} masses of 
neutrinos may be a result of SUSY. The latter is related to the notion of
R-parity violation.

The notion of R parity came about also  early in the history of SUSY\cite{Fy2}.
In those days, baryon and lepton number symmetries might look even better
than the standard model (SM) itself. R parity then seemed quite natural. However, 
global symmetries are since understood to be far less than sacred. The basic
theoretical building blocks of the SM are nothing more than the field spectrum 
and the gauge symmetries, while we have now strong evidence of nonzero 
neutrino masses that very likely cannot be fit into the pure Dirac mass picture. 
Why should one stick to R-parity conservation?

If the accidental symmetries of baryon number and lepton number in the SM 
are to be preserved in the supersymmetric SM, they would have to be added in 
by hand, {\it i.e.} imposed as extra global symmetries on the Lagrangian. 
R parity,  defined in terms of baryon number, lepton number, and spin as 
$R = (-1)^{3B+L+2S}$ does exactly that. This is, however, at the expense of 
making particles and superparticles having a categorically different quantum 
number. R parity is actually not the most effective discrete symmetry 
to control superparticle mediated proton decay resulted from having
both $B$ and $L$ violation\cite{pd}, but is most restrictive in terms of what 
is admitted in the Lagrangian, or the superpotential alone. Most importantly,
it separates the four extra neutral fermion states (called neutralinos) from the SM 
neutrinos, and keeps the latter massless. Giving up the {\it ad hoc} notion
of R parity, we naturally have massive neutrinos within the supersymmetric
SM without the need to add any extra superfields to the minimal spectrum. In this
way, one obtain massive neutrinos, from supersymmetry. More interestingly,
the theory also gives a rich range of lepton number violating phenomenology
from the same set of couplings that are responsible for the neutrino masses.

\section{The generic supersymmetric Standard Model}
The generic supersymmetric SM is a supersymmetrized SM with no extra symmetry,
R parity or otherwise, imposed. The model Lagrangian is simply the most general 
one constructed using the necessary (minimal) superfield spectrum, the gauge 
symmetries and renormalizability requirement, as well as the idea that SUSY is 
softly broken. One does expect some mechanism or symmetry to take care of the 
proton decay problem, which may also be naively taken as having a large 
enough suppression among the B violating couplings, from the phenomenological 
point of view. The lepton number and flavor violating couplings are good for 
incorporating the beyond SM properties of the neutrinos.
 
The most general renormalizable superpotential for the generic supersymmetric
SM can be written  as
\small\beqa
W \!\!\!\! &=& \!\!\!\!\varepsilon_{ab}\Big[ \mu_{\alpha}  \hat{H}_u^a \hat{L}_{\alpha}^b 
+ h_{ik}^u \hat{Q}_i^a   \hat{H}_{u}^b \hat{U}_k^{\scriptscriptstyle C}
+ \lambda_{\alpha jk}^{\!\prime}  \hat{L}_{\alpha}^a \hat{Q}_j^b
\hat{D}_k^{\scriptscriptstyle C} 
\nonumber \\
\!\!\!\!&& \!\! +\;
\frac{1}{2}\, \lambda_{\alpha \beta k}  \hat{L}_{\alpha}^a  
 \hat{L}_{\beta}^b \hat{E}_k^{\scriptscriptstyle C} \Big] + 
\frac{1}{2}\, \lambda_{ijk}^{\!\prime\prime}  
\hat{U}_i^{\scriptscriptstyle C} \hat{D}_j^{\scriptscriptstyle C}  
\hat{D}_k^{\scriptscriptstyle C}   ,
\eeqa\normalsize
where  $(a,b)$ are $SU(2)$ indices, $(i,j,k)$ are the usual family (flavor) 
indices, and $(\za, \zb)$ are extended flavor index going from $0$ to $3$.
In the limit where $\lambda_{ijk}, \lambda^{\!\prime}_{ijk},  
\lambda^{\!\prime\prime}_{ijk}$ and $\mu_{i}$  all vanish, 
one recovers the expression for the R-parity preserving case, 
with $\hat{L}_{0}$ identified as $\hat{H}_d$. Without R-parity imposed,
the latter is not {\it a priori} distinguishable from the $\hat{L}_{i}$'s.
Note that $\lambda$ is antisymmetric in the first two indices, as
required by  the $SU(2)$  product rules, as shown explicitly here with 
$\varepsilon_{\scriptscriptstyle 12} =-\varepsilon_{\scriptscriptstyle 21}=1$.
Similarly, $\lambda^{\!\prime\prime}$ is antisymmetric in the last two 
indices, from $SU(3)_{\scriptscriptstyle C}$. 

Doing phenomenological studies without specifying a choice of flavor bases is
ambiguous. It is like doing SM quark physics with 18 complex Yukawa couplings, 
instead of the 10 real physical parameters. As far as the SM itself is concerned, 
the extra 26 real parameters are simply redundant. There is simply no way to learn
about the 36 real parameters of Yukawa couplings for the quarks in some generic
flavor bases, so far as the SM is concerned. For instance, one can choose to
write the SM quark Yukawa couplings such that the down-quark Yukawa couplings
are diagonal, while the up-quark Yukawa coupling matrix is a product of (the
conjugate of) the CKM and the diagonal quark masses, and the leptonic Yukawa
couplings diagonal. Doing that is imposing no constraint or assumption onto the
model.  On the contrary, not fixing the flavor bases makes the connection between 
the parameters of the model and the phenomenological observables ambiguous.

In the case of the GSSM, the choice of flavor basis among the 4 $\hat{L}_\za$'s
is a particularly subtle issue, because of the fact that they are superfields the 
scalar parts of which could bear VEVs. A parameterization called the single-VEV
parameterization (SVP) has been advocated since Ref.\cite{ru1}.
The central idea is to pick a flavor basis such that only one among the 
$\hat{L}_\za$'s, designated as $\hat{L}_0$, bears a non-zero VEV. 
There is to say, the direction of the VEV, or the Higgs
field $H_d$, is singled out in the four dimensional vector space spanned by
the $\hat{L}_\za$'s. Explicitly, under the SVP, flavor bases are chosen such that :
1/  $\langle \hat{L}_i \rangle \equiv 0$, which implies
$\hat{L}_0 \equiv \hat{H}_d$;
2/  $y^{e}_{jk} (\equiv \lambda_{0jk} =-\lambda_{j0k})
=\frac{\sqrt{2}}{v_{\scriptscriptstyle 0}}\,{\rm diag}
\{m_{\scriptscriptstyle 1},
m_{\scriptscriptstyle 2},m_{\scriptscriptstyle 3}\}$;
3/ $y^{d}_{jk} (\equiv \lambda^{\!\prime}_{0jk})
= \frac{\sqrt{2}}{v_{\scriptscriptstyle 0}}\,{\rm diag}\{m_d,m_s,m_b\}$;
4/ $y^{u}_{ik}=\frac{\sqrt{2}}{v_{\scriptscriptstyle u}}\,
V_{\!\mbox{\tiny CKM}}^{\!\scriptscriptstyle T}\; {\rm diag}\{m_u,m_c,m_t\}$,
where $v_{\scriptscriptstyle 0}\equiv \sqrt{2} \, \langle \hat{L}_0 \rangle$
and $v_{\scriptscriptstyle u}\equiv \sqrt{2} \,
\langle \hat{H}_{u} \rangle$. 
The parameterization is optimal, apart from 
some minor redundancy in complex phases among the couplings. We 
simply assume all the admissible nonzero couplings within the SVP 
are generally complex. The big advantage of the SVP is that it gives 
the complete tree-level mass matrices of all the states (scalars 
and fermions) the simplest structure\cite{as8}.

\section{Neutrino Masses in the GSSM}
The GSSM has seven neutral fermions corresponding to the three
neutrinos and four, heavy, neutralinos. The heavy states are supposed
to be mainly gauginos and higgsinos, but there is now admissible 
mixings among all seven neutral electroweak states. 
In the case of small $\mu_i$'s of interest, it is convenient to use an
approximate seesaw block diagonalization to extract the effective
neutrino mass matrix. Note that the
effective neutrino mass here is actually written in a basis which
is approximately the mass eigenstate basis of the charged leptons,
{\it i.e.}, the basis is roughly $(\nu_e, \nu_\mu, \nu_\tau)$.
The tree-level result is very well-known\cite{ru6}. There have also been 
many papers devoted to the studies of radiatively generated neutrino masses
from R-parity violation. Here, we focus only on discussions under, essentially,
the current formulation\cite{as1,as5,as9,GH,ADL}.

The  neutral fermion mass matrix ${\mathcal M_{\!\scriptscriptstyle N}}$ can be 
written in the form of block submatrices:
\begin{equation} \label{mnu}
{\mathcal M_{\!\scriptscriptstyle N}} = \left( \begin{array}{cc}
              {\mathcal M}_n & \xi^{\!\scriptscriptstyle T} \\
              \xi & m_\nu \end{array}  \right ) \;,
\end{equation}
where ${\mathcal{M}}_n$ is the upper-left $4\times 4$ neutralino mass matrix, 
$\xi$ is the $3\times 4$ block, and $m_\nu$ is the lower-right 
$3\times 3$ neutrino block in the $7\times 7$ matrix. 
Starting with the generic formula
\beq \label{mpole}
{\mathcal M}(p^2) = {\mathcal M}(Q) + \Pi(p^2) -
{1\over 2}\, \left[\, {\mathcal M}(Q) \, \Sigma(p^2) + \Sigma(p^2) \, {\mathcal M}(Q)
\, \right] \; 
\eeq
casted in the electroweak basis, the effective neutrino mass matrix at 1-loop
level may be obtained as
\beqa
(m_\nu)^{\mbox{\tiny (1)}} &\simeq&
 - \xi \, {\mathcal M}_n^{\mbox{-}1} \, \xi^{\!\scriptscriptstyle  T}
+ \Pi_\nu + \Pi_\xi \,  {\mathcal M}_n^{\mbox{-}1} \,  \xi^{\!\scriptscriptstyle  T}
+ \xi \, {\mathcal M}_n^{\mbox{-}1} \, \Pi_\xi^{\!\scriptscriptstyle  T}
\nonumber\\ 
&+& {1\over 2} \, \Sigma_\nu \, \xi \,  {\mathcal M}_n^{\mbox{-}1} \,  \xi^{\!\scriptscriptstyle  T}
+ {1\over 2}  \, \xi \,  {\mathcal M}_n^{\mbox{-}1} \,  \xi^{\!\scriptscriptstyle  T}
\, \Sigma_{\nu}^{\!\scriptscriptstyle  T} 
+ \xi \, {\mathcal M}_n^{\mbox{-}1} \, \Pi_n \,
{\mathcal M}_n^{\mbox{-}1} \, \xi^{\!\scriptscriptstyle  T} \; ,
\label{m1loop}
\eeqa
where the $\Pi$'s and $\Sigma$'s denote two-point functions to be
evaluated at1-loop order. There are many pieces of contributions involving
various lepton number violating couplings. A neutrino mass term involves
violation of lepton number by two units, and basically any combination of
two lepton number violating couplings in the model contributes. Moreover,
to the extent that we are quite ignorant about the related phenomena, it is
dangerous to make any assumption about the relative strength of such
contributions, which is done quite often. We have indeed argued previously 
that the maximal mixing observed among neutrino flavors is likely to indicate
a flavor structure here very different from what we see among the other SM
fermions\cite{ru6}. Hence, it is rather necessary to check all the possible
contributions and have the general result ready. We have given exactly such
a listing\cite{as9}.

To keep within the length limit, we have to satisfy with only illustrating some
general feature of the detailed results. We are interesting in 1-loop two-point
functions with fermions and scalars, both charged or both neutral, running in 
the loop. An exact evaluation requires using mass eigenstates for the running
particles. We have given exact tree-level mass matrices for the five charged
fermions, seven neutral fermions, eight charged scalars, and ten neutral 
scalars\cite{as8}. Perturbational formulae for the elements of the 
diagonalization matrices are available\cite{as8}. The latter are very useful 
for an analytical understanding of the lepton flavor violating origin/structure of
each of the neutrino mass term. For instance, we have the charged loop
contribution
\beq \label{pinu}
%(m_\nu^o)_{ij} =
\Pi_{\nu_{ij}}^{\scriptscriptstyle C} = -
{\alpha_{\mbox{\tiny em}} \over 8 \pi \, 
\sin\!^2\theta_{\!\scriptscriptstyle W}} \;
% \frac{g_{\scriptscriptstyle 2}^2 }{32  \, \pi^2} \, \;
 {\mathcal C}_{\!\scriptscriptstyle inm}^{\scriptscriptstyle R^*} \,
 {\mathcal C}_{\!\scriptscriptstyle jnm}^{\scriptscriptstyle L} \;
{M}_{\!\scriptscriptstyle \chi^{\mbox{-}}_{n}} \, 
{\mathcal B}_{0}\!\!\left(p^2,{M}_{\!\scriptscriptstyle \chi^{\mbox{-}}_{n}}^2,
 {M_{\!\scriptscriptstyle \tilde{\ell}_{m}}^2} \right)\; ,
\eeq
where 
${\mathcal C}_{\!\scriptscriptstyle inm}^{\scriptscriptstyle R}$ and
${\mathcal C}_{\!\scriptscriptstyle jnm}^{\scriptscriptstyle L}$ are the effective
couplings of the $\nu_i$ and $\nu_j$ to the $m$-th charged scalar 
($\tilde{\ell}_m$) and $R$- and $L$-handed parts of the $n$-th charged 
fermion ($\chi_n^-$), respectively. Taking the $\lambda$-coupling term in
${\mathcal C}_{\!\scriptscriptstyle inm}^{\scriptscriptstyle R}$ and the gauge
coupling term in ${\mathcal C}_{\!\scriptscriptstyle jnm}^{\scriptscriptstyle L}$,
in particular, would resulted in a contribution proportional to 
\beq \label{mu-lam}
\frac{\lambda_{ikh}}{g_{\scriptscriptstyle 2}} \,
\mbox{\boldmath $V$}_{\!\!(h+2)n}^* \,  
{M}_{\!\scriptscriptstyle \chi^{\mbox{-}}_{n}} \, \mbox{\boldmath $U$}_{\!1n} \;
 D^{l}_{\!(k+2)m} \, {\mathcal D}^{l^*}_{\!(j+2)m}
\simeq
\frac{\lambda_{ijh}}{g_{\scriptscriptstyle 2}} \,
%\left( \frac{ R_{{\!\scriptscriptstyle R}_{21}}^*}{M_{c{\scriptscriptstyle 1}}} 
%+ \frac{ R_{{\!\scriptscriptstyle R}_{22}}^*}{M_{c{\scriptscriptstyle 2}}} 
%+ \frac{  \sqrt{2}\, M_{\!\scriptscriptstyle W} \cos\!\beta}{M_{\!\scriptscriptstyle 0}^2} \right) \,
\frac{m_h \, \mu_h}{M_{s}} \,.
%+ \frac{\lambda_{ikh}}{g_{\scriptscriptstyle 2}} \, 
%\frac{m_h \, \mu_h \, B_j \, B_k^*}{M_{s}^5 \, \cos^2\!\beta}
\eeq
%2nd term m=2, also m=j+2,k+2
We note here that the result is actually very sensitive to the $i\leftrightarrow j$ 
symmetrization. The dominant result in the expression above  is from the case with 
the $(j+2)$th charged scalar running in the loop. This is approximately the 
$\tilde{l}^{\!\!\mbox{ -}}_j$ slepton. The symmetrization and the fact that
$\lambda_{ijh}=-\lambda_{jih}$ suggest a perfect cancellation of the result in 
the limit of degenerate sleptons which correspond roughly to the   
$\tilde{l}^{\!\!\mbox{ -}}_i$ and $\tilde{l}^{\!\!\mbox{ -}}_j$ states\cite{as5,as9}.

\section{Beyond Neutrino Masses}
Our approach easily connect the neutrino mass results with a wide range of
other phenomenologies. An interesting class of such we have been studying
is electromagnetic dipole moments of the various fermions. Published results
are available for diagonal electric dipole moments of quark contributing
to neutron electric dipole moment\cite{as46} and  an example of transitional moments for the charged lepton giving rise to the 
$\mu \to e\,\gamma$ decay\cite{as7}.  One common important feature among
such dipole moment contributions is an interesting kind of contributions coming
from a combination of a bilinear and a trilinear lepton number violating
coupling --- a $\mu_i^*\lambda^\prime_{i\scriptscriptstyle 11}$ for $d$-quark
dipole moment and a $\mu_i^*\lambda_{i\scriptscriptstyle 21}$ for
$\mu \to e\,\gamma$. The constraints on the couplings we obtained from 
studies are actually close to comparable to neutrino mass constraints. Analyzes
of similar type of constraints from $b \to s\,\gamma$ and neutrino dipole
moment and radiative decays have are in progress.
 
We summarize results from our numerical study on the $BR<1.2\times10^{-11}$
experimental constraint on $\mu \to e\,\gamma$\cite{as7} in the following table
for your interest.
\vspace*{-.2in}
\begin{table}[h] \begin{center}
\begin{tabular}{|lr|}\hline 
\ \ 
$\frac{|{\mu_{\scriptscriptstyle 3}^*}\,{\lambda_{\scriptscriptstyle 321}}|}
{|\mu_{\scriptscriptstyle 0}|}\;, \;\;\;
\frac{|{\mu_{\scriptscriptstyle 1}^*}\,{\lambda_{\scriptscriptstyle 121}}|}
{|\mu_{\scriptscriptstyle 0}|}\;, \;\;\;
\frac{|{\mu_{\scriptscriptstyle 3}}\,{\lambda_{\scriptscriptstyle 312}^*}|}
{|\mu_{\scriptscriptstyle 0}|}\;, \;\;\; 
\mbox{or} \;\;\;
\frac{|{\mu_{\scriptscriptstyle 2}}\,{\lambda_{\scriptscriptstyle 212}^*}|}
{|\mu_{\scriptscriptstyle 0}|}\; \;\;\;$ & 
$< 1.5 \times 10^{-7}$ \ \ 
\\ \ \
$\frac{|\mu_{\scriptscriptstyle 1}^*\, \mu_{\scriptscriptstyle 2}|}
{|\mu_{\scriptscriptstyle 0}|^2}$	&	$ < 0.53 \times 10 ^{-4}$ \ \
\\ \ \
$|\lambda_{\scriptscriptstyle 321} \lambda^*_{\scriptscriptstyle 131}|\;, \;\;\;
|\lambda_{\scriptscriptstyle 322} \lambda^*_{\scriptscriptstyle 132}|\;, \;\;\; 
\mbox{or} \;\;\;
|\lambda_{\scriptscriptstyle 323} \lambda^*_{\scriptscriptstyle 133}|$      
& $<2.2 \times 10^{-4}$ \ \ 
\\ 
\ \
$|\lambda^*_{\scriptscriptstyle 132} \lambda_{\scriptscriptstyle 131}|\;, \;\;\;
|\lambda^*_{\scriptscriptstyle 122} \lambda_{\scriptscriptstyle 121}|\;, \;\;\; 
\mbox{or} \;\;\;
|\lambda^*_{\scriptscriptstyle 232} \lambda_{\scriptscriptstyle 231}|$      
& $<1.1 \times 10^{-4}$ \ \ 
\\ 
\ \  
$\frac{|B_{3}^*\,\lambda_{\scriptscriptstyle 321}|}{|\mu_{\scriptscriptstyle 0}|^2}
\;, \;\;\;
\frac{|B_{1}^*\,\lambda_{\scriptscriptstyle 121}|}{|\mu_{\scriptscriptstyle 0}|^2}
\;, \;\;\;
\frac{|B_{3}\,\lambda_{\scriptscriptstyle 312}^*|}{|\mu_{\scriptscriptstyle 0}|^2}
\;, \;\;\; \mbox{or} \;\;\;
\frac{|B_{2}\,\lambda_{\scriptscriptstyle 211}^*|}{|\mu_{\scriptscriptstyle 0}|^2}$
&  $<2.0\times 10^{-3}$ \ \ 
\\ 
\ \ 
$\frac{|B_1^* \, \mu_{\scriptscriptstyle 2}|}{|\mu_{\scriptscriptstyle 0}|^3}$
 & $< 1.1\times 10^{-5}$  \ \ 
\\ \hline
\end{tabular}\end{center}
\end{table}

\vspace*{-.2in}
\noindent
The numbers are based inputs as given by

\begin{table}
\begin{center}
%\vspace*{.2in}
\begin{tabular}{|cccc|}
\hline
$M_{\scriptscriptstyle 1}$ (GeV) & $M_{\scriptscriptstyle 2}$ (GeV)  & $\mu_{\scriptscriptstyle 0}$ (GeV)  & $\tan\!\beta$ \\
\hline 
100  & 200 & 100 & 10 \\
\hline
\hline
$\tilde{m}^2_{\!{\scriptscriptstyle L}}$ ($10^4$ GeV$^{2}$) &
$\tilde{m}^2_{\!{\scriptscriptstyle E}}$ ($10^4$ GeV$^{2}$) & 
$A_e$ (GeV) & \\
\hline
diag$\{2,1,1,1\}$ &
diag$\{1,1,1\}$ & 100 & \\
\hline
\end{tabular}
\end{center}
\end{table}

\section{Concluding Remarks}
Supersymmetry could be considered a source of neutrino masses and other beyond
SM properties of neutrinos. Promoting the field multiplet spectrum of SM to superfields
gives naturally lepton number and flavor violating couplings admissible by the
gauge interactions. In that sense, the result generic supersymmetric SM is the
simplest supersymmetric model incorporating neutrino masses. Other alternatives
require extra superfields beyond the minimal spectrum, and usually also {\it ad hoc}
global symmetries, in some case with specifically assumed symmetry breaking
patterns. Another attractive feature of the generic supersymmetric SM is that the same
set of couplings giving the neutrino masses also give rise to a width range of 
lepton number and flavor violating interactions. There is then correlation between
the neutrino masses and other (collider) phenomenologies to be explored. 
Our formulation, called single-VEV parameterization, has been demonstrated to give
a very effective framework to simplify any analytical studies of the model, making 
the task within easy reach. The whole discussion here is based on a purely
phenomenological perspective. We are suggesting studying all the experimental
constraints we could obtained on the set of couplings without theoretical bias. 
The hope to that we could eventually find some pattern among them and learn about
the problem of the flavor structure among them. The lesson we learned so far ,from
neutrino masses and mixings, is that the usually hierarchical flavor structure 
established among the Yukawa couplings of the quarks and charged leptons simply
does not apply here. However, the lepton number violating couplings revealed 
through neutrino properties and otherwise may one day help to shed a light on 
the general flavor problem.

%\newpage

\end{document}